\documentclass[a4,article,nofootinbib]{revtex4}
\usepackage{graphicx}
\usepackage{epstopdf, epsfig}
\usepackage{amssymb} 
\usepackage{enumerate}
\usepackage{amsmath} 

\usepackage{hyperref} 

\usepackage{stmaryrd}

\usepackage{upgreek}

\newcommand{\muSI}{\upmu_0} 
\newcommand{\epsilonSI}{\upepsilon_0} 

\newcommand{\bs}[1]{\boldsymbol{#1}} 
\renewcommand{\d}{\mathrm{d}} 

\newcommand{\vdot}{\mbox{\boldmath\(\cdot\)}}
\newcommand{\cross}{\mbox{\boldmath\(\times\)} }
\newcommand{\grad}{\mbox{\boldmath\(\nabla\)} }
\newcommand{\divv}{\mbox{\boldmath\(\nabla\vdot\)} }
\newcommand{\curl}{\mbox{\boldmath\(\nabla\times\)} }

\newcommand{\const}{{\mathrm{const}}}

\newcommand{\mc}{\mathcal}

\newcommand{\bol}{\boldsymbol}

\newcommand{\abs}[1]{\left\lvert{#1}\right\rvert}

\newcommand{\lr}[1]{\left({#1}\right)}

\newcommand{\pa}{\partial}

\begin{document}
\title{Relaxation of Compressible Euler Flow in a Toroidal Domain}
\author{N. Sato}
\affiliation{Graduate School of Frontier Sciences, The University of Tokyo,
Kashiwa, Chiba 277-8561, Japan}
\author{R. L. Dewar}
\affiliation{Centre for Plasmas and Fluids, Research School of Physics and Engineering, The Australian National University,
 Canberra, ACT 2601, Australia}
\date{\today}



\begin{abstract}
It is shown that the \emph{universal} steady Euler flow field, independent of boundary shape or symmetry, in a toroidal domain with fixed boundary obeys a nonlinear Beltrami equation, with the nonlinearity arising from a Boltzmann-like, velocity-dependent factor. Moreover, this is a \emph{relaxed} velocity field, in the sense that it extremizes the total kinetic energy in the domain under free variations of the velocity field, constrained only by tangential velocity and vorticity boundary conditions and conservation of total fluid helicity and entropy. This is analogous to Woltjer-Taylor relaxation of plasma magnetic field to a stationary state. However, unlike the magnetic field case, attempting to derive slow, quasi-relaxed \emph{dynamics} from Hamilton's action principle, with constant total fluid helicity as a constraint, fails to agree, in the static limit, with the nonlinear Beltrami solution of the Euler equations. Nevertheless, an action principle that gives a quasi-relaxed dynamics that \emph{does} agree can be formulated, by introducing a potential representation of the velocity field and defining an analogue of the magnetic helicity as a new constraint. A Hamiltonian form of quasi-relaxed fluid dynamics is also given.
\end{abstract}


\maketitle

\section{Introduction}\label{sec:1} 

A recently proposed reformulation of magnetohydrodynamics (MHD), multi-region relaxed MHD (MRxMHD), \cite{Hudson12,Dewar15, Dewar_Tuen_Hole_17, Dewar_Hudson_Bhattacharjee_Yoshida_17}, appropriate for modelling the dynamics and statics of toroidal plasmas with no continuous symmetry, couples magnetic field and fluid only at flexible boundaries partitioning the system into discrete toroidal sub-regions.  Within such a sub-region, $\Omega$ say, the magnetic field is a linear Beltrami field, which is force-free and thus decoupled from the plasma, which is a compressible ideal (inviscid or \emph{Euler}) fluid with an isothermal equation of state. The dynamics of this thermally relaxed Euler fluid was derived by \cite{Dewar15} from a Lagrangian, but a representation of Euler flow dynamics in Hamiltonian form, using non-canonical Poisson brackets, has been given by \cite{Morrison_98}. We shall mainly use a Lagrangian approach, but a Hamiltonian representation of our final result will be given at the end of the paper.

As the context of this work is toroidal plasma confinement, we take $\Omega$ to be a \emph{toroid}, by which we mean a three-dimensional (3-D) volume bounded by one or two 2-tori, the latter case being that of an \emph{annular} toroid, \cite{Dewar15}. Thus the flow is \emph{recirculating}, which motivates us to revisit the problem, \cite{Grimshaw69}, of steady recirculating Euler flows, and to derive the slow dynamics of almost steady flows. Generalization of the results to multiple domains being straightforward, our attention will be limited to a single relaxation region in order to simplify the exposition. Thus, throughout the paper, the boundary $\partial\Omega$ is taken to be some prescribed function of space and time.

The present paper is organized as follows. 
In Section~\ref{sec:2} we discuss the concept of relaxation, in particular in the context of  magnetohydrodynamics (MHD) and fluid mechanics. In section \ref{sec:3}, we derive, directly from the Euler equations, a steady flow equation 
that is valid for arbitrarily chaotic flow lines and is therefore valid for \emph{any} boundary shape, symmetric or non-symmetric. We term this most robust flow, satisfying a nonlinear Beltrami equation, the \emph{universal flow}. In section~\ref{sec:4} we also identify the universal flow as a \emph{relaxed} flow, in the sense that it makes the kinetic plus thermodynamic internal energy stationary under variations satisfying only the macroscopic constraints of total mass, entropy and fluid helicity conservation. 

In section \ref{sec:5}, we formulate an \emph{action} principle for the slow dynamics of the relaxed fluid as it evolves, on a timescale greater than or equal to a relaxation time, from small initial departures from a steady state or responds to slowly time-varying boundary perturbations. We then obtain the associated Euler-Lagrange equations
and show they are consistent, in the time-independent limit, with the relaxed state found directly, and with an \emph{energy} principle, in section~\ref{sec:4}.  However, we show that this consistency between the energy and action principles is obtained only if a Helmholtz potential representation for the flow velocity is used in the action principle, varying these potentials under a \emph{new} helicity-like constraint written in terms of the vector potential rather than the velocity itself.

In section \ref{sec:6} we discuss the physical properties of the relaxed fluid and determine their relationship with the Euler fluid equations. In section \ref{sec:7} we address the question of stability of the stationary solution to the relaxed Euler equations. Conclusions are drawn in section \ref{sec:8}.
Finally, an equivalent Hamiltonian formulation of  the relaxed Euler fluid is given in Appendix~\ref{sec:HamForm},  and, in Appendix~\ref{sec:OhmLaw} a development is given of a method to reduce the violation of the ideal Ohm's Law implied by the decoupling of $\bs{v}$ and $\bs{B}$ in MRxMHD equilibria by an optimum choice of electric field. This reduces the dynamo (assumed to arise from ongoing mesoscale turbulence) needed to explain the self-organizing effect of relaxation, helps sets the current paper in the broader MHD context, and provides an example that marries magnetic field and fluid relaxation.

\section{Relaxation}\label{sec:2} 

The \emph{relaxation} of a system, initially not in thermodynamic and mechanical equilibrium, is any process that drives it toward a steady \emph{relaxed state} compatible with any constraints imposed on the system. For an isolated system, this state is a \emph{maximum entropy} state. However, if the system in question is not thermodynamically isolated but is instead a subsystem [in the abstract sense discussed by \cite{Penfield_Haus_67}] of a larger system that absorbs the kinetic energy converted to heat or radiation during the relaxation process, the relaxation leads to \emph{minimum energy} state. (As a simple example, consider the settling of the centre of mass of a rolling ball toward the lowest point accessible to it --- the dissipated energy is absorbed by the environment and the ball itself.) While, for some purposes, these two definitions of a relaxed state are equivalent, \cite{Finn_Antonsen_83}, we follow the majority of the MHD literature in adopting the energy approach because it fits more naturally into our dynamical formulation. It should also be mentioned that, in a Hamiltonian formulation, existence of a minimum energy state arises formally as a necessary and sufficient condition for stability using the Energy-Casimir (Lyapunov) method, see e.g. the review by \cite{Morrison_98}.

Relaxed states play a fundamental role in the description of physical systems in many areas of the natural sciences, 
ranging from those fields where the arguments of equilibrium statistical mechanics apply, to the intrinsically non-equilibrium fluid and plasma problems encountered in both terrestrial and astrophysical areas. 
The accessibility of such relaxed states is 
related to the presence of dissipative mechanisms, such as viscosity in a fluid or resistivity in a magnetized plasma, that violate the ideal, non-dissipative picture of a system that often provides the most tractable starting point for describing its physical behaviour. 
It is thus desirable to find models that retain (or increase) the attractive simplicity of the ideal picture yet capture relaxation phenomena. 

Regarding the constraints alluded to above, we distinguish two classes, \emph{microscopic} and \emph{macroscopic}, the former referring to individual fluid elements and the latter referring to the fluid throughout the domain $\Omega$. 
The existence of constraints, 
on a timescale long compared with the time scale of the relaxation process, 
prevents the achievement of 
total 
thermodynamic equilibrium and imparts nontrivial structures, such as flows and vortices to the system. 


For a general review of relaxation, in physical contexts of the kind we consider in this paper, see \citet{Dewar08}. A specific and much cited example of a relaxed system is the relaxation of a magnetic field to a linear force-free state, introduced in the context of astrophysical plasmas by \citet{Chandra58} and \citet{Woltjer58}, and developed extensively in a laboratory context by \citet{Taylor74,Taylor86}, leading it to become known as \emph{Taylor relaxation}. 

The concept of Taylor relaxation is 
conceptually  
related to 
statistical mechanical 
relaxation by the notion of coarse graining (local averaging, or smoothing).
In Taylor relaxation there is an implicit ergodic assumption that field lines sample the whole
configuration space within $\Omega$, due to small-scale turbulence and/or other effects 
leading to field-line chaos, which is assumed to break the microscopic frozen-in-flux constraint of ideal MHD (IMHD). Even if, as is typical, there are invariant tori and islands blocking complete chaos at any particular time in a given shot, as long as there is unpredictability we can also borrow the concept of ensemble averaging from statistical mechanics to weaken the need to assume complete field-line chaos.

The Taylor relaxed state is a Beltrami field, i.e. a vector field $\boldsymbol{B}$ completely aligned to its own curl, $\boldsymbol{B}=\mu^{-1}\curl\boldsymbol{B}$, and is linear because $\mu$ is constant. Taylor's physical argument concerned the minimization of the magnetic energy (assumed to be the dominant contribution to the energy) due to dissipation occurring during the relaxation toward the quiescent state. 
The only IMHD invariant retained during this minimization [apart from the constancy of macroscopic magnetic fluxes, which is a boundary constraint, \citet{Dewar_Hudson_Bhattacharjee_Yoshida_17}] was assumed to be the macroscopic constraint of constant magnetic helicity, \citet{Woltjer58}, leading to a very simple variational problem.  
This assumption turns out to be
valid when the relaxation is caused by small scale turbulence and the robustness of the magnetic helicity is ensured by its rate of change, which becomes negligible if compared with the rate of energy dissipation \citep[see][]{Taylor86,Jensen84,Ito96,Qin12}.

The effectiveness of Taylor's theory in describing relaxed plasma states is known to go beyond those systems that satisfy the prescriptions of Taylor relaxation \citep[see][]{Qin12}. The physical reason is that the Taylor linear Beltrami state admits a dual interpretation in statistical mechanical terms according to which it represents a maximum entropy state or an extremum of the entropy production rate \citep[see][]{Finn_Antonsen_83,Minardi05,Hameiri87,Dewar08}.

The Taylor state applies to the magnetic field alone and does not make physical predictions on
the form of the velocity field $\boldsymbol{v}$ of the relaxed fluid. The form of $\boldsymbol{v}$ is, however, 
a fundamental feature in the design of any magnetic confinement device and 
the main aspect of interest from the fluid mechanics standpoint. 
One may argue that, since  
the magnetohydrostatic and (incompressible) steady flow equations are isomorphic, and that the flow helicity parameter\footnote{Note: The insertion of a factor $1/2$ in defining $K^{\rm fl}$ will be found algebraically convenient as it matches the $1/2$ in the kinetic energy, but it makes it \emph{half} the flow helicity $\mathcal{H}$ as normally defined, see e.g. \cite{Moffatt14}.}
\begin{equation}
K^{\rm fl}_{\Omega}=\int_{\Omega}{\frac{\boldsymbol{v}\vdot\boldsymbol{\omega}}{2}} \,dV\label{Kfl}
\end{equation}
is an invariant of the Euler equations for the motion of a barotropic inviscid fluid subjected to conservative body forces and such that the vorticity $\boldsymbol{\omega}=\curl\boldsymbol{v}$ has no component normal to the bounding surface \citep[see][]{Moffatt69,Moffatt92,Moffatt14}, a Taylor state analogous to that for the magnetic field should exist also for the velocity field $\boldsymbol{v}$ with the kinetic energy
\begin{equation}
	W_{\Omega}^{v}=\int_{\Omega}{\rho\frac{v^{2}}{2}} \,dV
\end{equation}
as target functional. Here $\rho$ is the fluid density, $dV$ the infinitesimal volume element, and $\Omega$ the volume occupied by the fluid.
As in the Taylor problem, we assume (see \ref{sec:1}) 
the domain $\Omega$ to be topologically toroidal, so that flow lines \emph{recirculate} indefinitely.

However, both theory and experimental evidence suggest that the analogy between $\boldsymbol{B}$ and $\boldsymbol{v}$ is 
not exact due to 
the problem of stability. Specifically, it is found that, while magnetostatic solutions for the magnetic field $\boldsymbol{B}$ may be stable against perturbations (including the Taylor state, provided $|\mu|$ is below the first Beltrami eigenvalue), steady solutions of the Euler equations (including the Beltrami state for $\boldsymbol{v}$) do not satisfy the analogous stability criterion with respect to non-steady perturbations obeying the dynamical Euler equations \citep[see][]{Moffatt14,Moffatt85,Moffatt86}. 

That being said, the fluid stability situation is not clear cut as the criterion for stability in question is only sufficient, not necessary, so one needs to consider the specific physical context carefully. We note in particular that it is easy to find stable two-dimensional (2-D) flows, so geometry matters, and that the context we are considering is (see Section~\ref{sec:1}) flow in a toroid of which the average radial dimension is considerably less than the average minor radius of the plasma and much less than the poloidal and toroidal periodicity lengths. It is thus reasonable to conjecture that long-wavelength relaxed flows in $\Omega$ can be regarded as quasi-2-D and stable. 
Hence, quasi-2-D turbulence \citep[see][]{Shi} provides a mechanism to generate a self-organized and relaxed fluid state \citep[see][]{Hasegawa}, which is expected to be `dynamical' as long as the corresponding time scale is slower than the relaxation time of the dissipative process.
While departures from a relaxed state may excite instabilities of short (mesocopic) wavelength, these will either participate in a Kolmogorov cascade to the microscopic dissipation range or in an inverse cascade to the macroscopic scale, thus giving a physical path to re-form the quasi-2-D relaxed state. [Bidirectional cascades have been studied in other contexts, see e.g. \cite{Korotkevich_08}.]

In addition, it is worth
noting 
that the presence of a magnetic field can be a source of stability:
as shown 
by \citet{Moffatt86}, 
if $\bs{B}$ and $\bs{v}$ are described in terms
of the same Beltrami equilibrium, they minimize the energy functional.   
We refer the reader to \citep{Hussain86,Rogers87,Wallace92} for some experimental studies
on stationary fluid equilibria. 
While not directly relevant to the present paper as the fluid in $\Omega$ is assumed to be completely decoupled from any magnetic field and $\partial\Omega$ is prescribed,  
this effect is briefly proposed in Section~\ref{sec:7} as a possible mechanism for stabilizing $\partial\Omega$ if it is instead a current/vortex sheet between adjacent regions in the MRxMHD model cited in Section~\ref{sec:1} and discussed further in Appendix~\ref{sec:OhmLaw}. 

Furthermore, there is good evidence that the flow helicity $K^{\rm fl}_{\Omega}$ is still a relevant topological invariant of the Euler flow, even in the presence of viscosity \citep[see][]{Scheeler14} which causes transfer, rather than dissipation, of helicity from larger to smaller scales
in a Kolmogorov cascade, or from 
smaller to larger scales in an inverse cascade. 
Assuming the majority of the helicity is in the longer-wavelength inverse cascade range (so the helicity eventually lost in the dissipation range is negligible), we can thus take $K^{\rm fl}_{\Omega}$ to be invariant even during dynamical relaxation, when turbulence is excited to maintain a close-to-relaxed ``condensate,'' \cite{Korotkevich_08}, during slow time changes in the boundary.  

Concerning the construction of a relaxed fluid state, most of the efforts made so far pertain to generalizations of Taylor's relaxation theory. In \citet{Steinhauer97} a steady flow results by considering a multi-fluid system (ions and electrons) where the constraints for the variational problem that minimizes the magnetic energy are the canonical helicities $K^{\rm c,\alpha}_{\Omega}=\int_{\Omega}{\boldsymbol{P}_{\alpha}\vdot\curl\boldsymbol{P}_{\alpha}} dV$, with $\boldsymbol{P}_{\alpha}=m_{\alpha}\boldsymbol{v}_{\alpha}+q_{\alpha}\boldsymbol{A}$ and where $m_{\alpha}$, $\boldsymbol{v}_{\alpha}$, and $q_{\alpha}$ are mass, velocity, and charge of specie $\alpha$. 
An equivalent multi-fluid approach \citep[see][]{Mahajan98} leads to a double Beltrami steady state in both $\boldsymbol{B}$ and $\boldsymbol{v}$. 
A steady state with flow is also derived in \citet{Dennis14} within the framework of IMHD by introducing cross helicity and toroidal angular momentum as additional global invariants [see also \citet{Lingam16} for a generalization to Hall MHD]. 
Important progress was made in \citet{Dewar15}, where a dynamical (time dependent) relaxed state for the flow velocity $\boldsymbol{v}$ is obtained through Hamilton's principle of stationary action \cite[see][]{Penfield66a,Penfield66b}. 
The resulting dynamical equations 
respect 
a subset of IMHD 
constraints 
and are referred to as MRxMHD (multi-region relaxed magnetohydrodynamics). The basic idea behind this generalization of Taylor's relaxation theory is that the relaxed state is made accessible by freeing the weaker constraints (specifically, the microscopic freezing of entropy and magnetic flux in each fluid element) and by enforcing global invariants (global entropy, magnetic flux, and helicity).
Mathematically this is achieved through the MHD Lagrangian density \citep[see][]{Newcomb62,Dewar70,Dewar78}, 
but now with  
pressure and magnetic field treated as independent variables in the variational problem,
rather than being holonomically constrained to fluid displacements as done in the IMHD work of \cite{Dewar70}.  
The term multi-region refers to the fact that relaxation is carried out over different domains separated by current sheets that enable different Beltrami states across the magneto-fluid. 
 
The aim of the present paper is to 
extend 
the relaxation theory developed in \citet{Dewar15} 
to include the concepts of static and dynamic relaxation of the mass flow velocity field $\bs{v}$. 
This is achieved by freeing the 
velocity field from the Lagrangian constraint 
\begin{equation}
\frac{d\boldsymbol{x}}{dt}=\boldsymbol{v},\label{dxdt}
\end{equation}
linking $\bs{v}$ to the motion of a fluid element at  $\bs{x}(\bs{x}_0,t)$, where $\bs{x}_{0}$ is the position of the fluid element at initial time $t_0$. 
The role of the constraint \eqref{dxdt}, which implies that initial conditions $\bs{x}_{0}\left(\bs{x},t\right)$ are always preserved by $\bs{v}$, in the derivation of the Euler-Lagrange equations for general fluid dynamics is outlined in \citet{Penfield66a,Penfield66b,Yoshida12}. 
The removal of this constraint is necessitated by the subtle change in meaning of the Eulerian flow field $\bs{v}(\bs{x},t)$, which now describes a \emph{mean flow} \citep[see e.g.][]{Buhler09}, in which mesoscopic turbulent fluctations are averaged over, and information on initial positions is rapidly lost due to flow-line chaos. 

In order to free the velocity, we treat its variations as independent from those in position and enforce the global flow helicity invariant \eqref{Kfl}.
As it will be shown in the following sections, this variation needs careful treatment in order to ensure the local
preservation of mass.
Physically, the operation described above 
is equivalent (in the dual entropy/energy way explained earlier in this section) to  
introducing the effect of viscosity, which drives $\boldsymbol{v}$ toward the maximum entropy state. 
The result of this construction is the \emph{relaxed Euler fluid}.
Here, the fluid flow $\bs{v}$ is described by a time-dependent non-linear Beltrami field with a non-constant multiplier controlled by the fluid density $\rho$, and a displacement term $\bs{d}$ as $\bs{\omega}\propto\rho\bs{v}+\bs{d}$. This non-steady 
quasi-equilibrium 
can account for time dependent behavior and misalignment between $\boldsymbol{v}$ and $\boldsymbol{\omega}$. 
As we will show, the Beltrami equilibrium of the relaxed Euler equations is stable under appropriate conditions. 


\section{General steady flow solution}\label{sec:3}
 
The compressible, inviscid ideal (Euler) fluid  
is described by the following set of equations for fluid mass density $\rho$, 
flow velocity $\boldsymbol{v}$, and pressure $P$: 
\begin{subequations}\label{Euler}
\begin{align}
\frac{\partial\rho}{\partial t}&=-\divv\left(\rho\boldsymbol{v}\right),\label{continuity}\\
\frac{d\boldsymbol{v}}{dt}&=-\rho^{-1}\grad P,\label{momentum}
\end{align}
\end{subequations} 
\eqref{continuity} is the continuity equation, expressing local mass conservation.
\eqref{momentum} is the momentum equation. For the time being, we shall further assume the barotropic equation of state:
\begin{equation}
P=\tau\rho\label{state}.
\end{equation}
We will see later on how this equation naturally follows from the variational relaxation principle.
The quantity $\tau = C_{{\rm s}}^2$, is identifiable physically as the temperature in energy units, divided by ion mass.
$C_{\rm s}$ is the speed of sound in the appropriate units. 
Setting time derivatives to zero in \eqref{continuity} and \eqref{momentum}, we have the mass conservation equation
\begin{equation}\label{eq:mass}
	\divv(\rho\boldsymbol{v}) = \boldsymbol{v}\vdot\grad\rho + \rho\divv\boldsymbol{v} = 0
\end{equation}
and the centrifugal acceleration equation
\begin{equation}\label{eq:momentum1}
	\boldsymbol{v}\vdot\grad\boldsymbol{v} = -\frac{\tau \grad\rho}{\rho} \;,
\end{equation}
which can be written
\begin{equation}\label{eq:momentum2}
	(\curl\boldsymbol{v})\cross\boldsymbol{v} = -\grad\left(\frac{v^2}{2} + \tau\ln\frac{\rho}{\rho_0}\right) \;,
\end{equation}
where $\rho_0$ is a constant reference density, at this stage arbitrary.
Dotting both sides with $\boldsymbol{v}$ gives a form of the Bernoulli equation for compressible flows
\begin{equation}\label{eq:Bernoulli1}
	\boldsymbol{v}\vdot\grad\left(\frac{v^2}{2} + \tau\ln\frac{\rho}{\rho_0}\right) = 0 \;.
\end{equation}
This equation implies $v^2/2 + \tau\ln(\rho/\rho_0) = \const$ along a streamline.
Therefore, the solution depends on whether or not the streamline flow is integrable. 
In the integrable case the flow foliates space with invariant tori, which are level surfaces of the smoothly varying spatial function $v^2/2 + \tau\ln(\rho/\rho_0)$. However, three-dimensional flows are not generically integrable, 
and thus the only solution that is not hopelessly complicated is such that $v^2/2 + \tau\ln(\rho/\rho_0)$ is constant throughout $\Omega$, corresponding to the ``worst case'' confinement scenario where a single streamline fills $\Omega$ ergodically (analogous reasoning in \citet{Hudson12} leads to the adoption of linear force-free magnetic fields as the basic building blocks of MRxMHD). By appropriate choice of $\rho_0$, we can make this constant zero,
\begin{equation}\label{eq:Bernoulli2}
	\frac{v^2}{2} + \tau\ln\frac{\rho}{\rho_0} = 0 \;.
\end{equation}
Introducing the Boltzmann factor $\exp(-v^2/2\tau)$ and using \eqref{eq:Bernoulli2} we find explicit expressions for $\rho$ and $P$,
\begin{subequations}\label{eq:Bernoulli3}
\begin{align}
	\rho &= \rho_0\exp\left(-\frac{v^2}{2\tau}\right),\\
	     P &= P_0\exp\left(-\frac{v^2}{2\tau}\right)\;,
\end{align}
\end{subequations}
where $P_0 = \tau \rho_0$.

With the solution \eqref{eq:Bernoulli3}, \eqref{eq:momentum2} becomes $(\curl\boldsymbol{v})\cross\boldsymbol{v} = 0$, i.e. the \emph{velocity and vorticity are aligned} (cf. the force-free magnetic field of the Taylor state)
, giving a Beltrami-like equation for $\boldsymbol{v}$.
\begin{equation}\label{eq:vBeltrami}
	\curl\boldsymbol{v} = \alpha\boldsymbol{v} \;.
\end{equation}
Taking the divergence of \eqref{eq:vBeltrami} gives
\begin{equation}\label{eq:divvBeltrami}
	\boldsymbol{v}\vdot\grad\alpha + \alpha\divv\boldsymbol{v} = 0\;,
\end{equation}
which, being isomorphous to \eqref{eq:mass}, implies $\alpha/\rho = \const$ throughout $\Omega$, and, 
from \eqref{eq:Bernoulli3}, that there exists a constant $\alpha_0$ such that
\begin{equation}\label{eq:BeltramiBernoulli}
	\alpha = \alpha_0\exp\left(-\frac{v^2}{2\tau}\right) \;,
\end{equation}
which is constant only if $v^2$ is constant or if $\tau \to \infty$, i.e. in the deep subsonic limit $v/C_{{\rm s}} \to 0$. Related results have been found by \citet{Grimshaw69} in the incompressible case.

Thus \eqref{eq:vBeltrami} is the Beltrami equation with non constant multiplier
\begin{equation}\label{eq:vBeltramiNL}
	\curl\boldsymbol{v} = \alpha_0\exp\left(-\frac{v^2}{2\tau}\right)\boldsymbol{v} \;,
\end{equation}
to be solved under the tangential free-slip boundary condition $\boldsymbol{n}\vdot\boldsymbol{v} = 0$ on the boundary $\partial\Omega$, $\boldsymbol{n}$ being the (outward) unit normal on $\partial\Omega$.
We expect \eqref{eq:vBeltramiNL} to be the steady flow limit of the  relaxed state 
resulting by freeing the constraint \eqref{dxdt}.


\section{Energy relaxation principle}\label{sec:4}

Before considering a dynamical formulation, it is worth noticing that
the stationary state \eqref{eq:vBeltramiNL} can be obtained in terms of the energy principle:
\begin{equation}
\delta\lr{\mc{H}_{\Omega}-\frac{\rho_{0}}{\alpha_{0}} K^{\rm fl}_{\Omega}-\tau S_{\Omega}}=0.
\end{equation}
Here, the independent variables are $\bol{v}$, $\rho$ and $P$, $\rho_{0}/\alpha_{0}$ and $\tau$ act as Lagrange multipliers, $\mc{H}_{\Omega}$ is the fluid energy:
\begin{equation}
\mc{H}_{\Omega}=\int_{\Omega}{\lr{\rho\frac{v^{2}}{2}+\frac{P}{\gamma-1}}}\,dV,\label{Energy}
\end{equation}
and $S_{\Omega}$ the global entropy \citep[see appendix A of][]{Dewar15}: 
\begin{equation}
S_{\Omega}=\int_{\Omega}\frac{\rho}{\gamma-1}\log\left(\kappa\frac{P}{\rho^{\gamma}}\right)\,dV.\label{S}
\end{equation}
In \eqref{S} we introduced the constants $\gamma$ and $\kappa$.
$\gamma$ is the constant of the adiabatic equation of state $P\rho^{-\gamma}={\rm const}$, with $P$ the pressure, while $\kappa$ ensures that the argument of the logarithm is dimensionless. 

\section{Action relaxation principle}\label{sec:5}

\subsection{Independent fields}

The relaxed state for the fluid flow follows by treating $\boldsymbol{v}$ as an independent variable 
in the application of Hamilton's principle of stationary action.
Indeed, the turbulent velocity $\widetilde{\boldsymbol{v}}$ responsible for the relaxation decouples the (mean) fluid velocity $\boldsymbol{v}$ from the fluid position $\boldsymbol{x}$, since $\boldsymbol{v}+\widetilde{\boldsymbol{v}}=\frac{d\boldsymbol{x}}{dt}$.
However, the velocity field $\boldsymbol{v}$ is not completely independent from variations in the fluid density, 
due to the holonomic constraint imposed by the continuity equation \ref{continuity}. 
In order to enforce such constraint, we need to determine the actual independent fields that can be
varied to extremize the action with a procedure analogous to that used in electrodynamics to ensure the preservation of charge.
First, we observe that the continuity equation states that the four dimensional
vector field:
\begin{equation}
\mathcal{V}=\rho\partial_{t}+\rho\boldsymbol{v},\label{V}
\end{equation}  
is divergence free. In this notation $\partial_{t}$ is the unit vector along the $t$-axis.
Therefore, there is a $3$-form $\omega$ such that
\begin{equation}
\omega=d\beta,\label{alpha}
\end{equation}
where $\beta$ is a $2$-form and
\begin{equation}
\omega=i_{\mathcal{V}}vol^{4},\label{alpha2}
\end{equation}
with $vol^{4}=dt \wedge dx \wedge dy \wedge dz$. 
Note that \eqref{V} is divergence free because from \eqref{alpha} and \eqref{alpha2} we have $d\omega=\left(\partial_{t}\rho+\divv\rho\boldsymbol{v}\right)vol^{4}=0$.
The general form of $\beta$ is $\beta=\frac{1}{2}\beta_{\mu\nu}dx^{\mu}\wedge dx^{\nu}$, which implies that:
\begin{equation}
\ast d\beta = \left(\divv\boldsymbol{w}\right)dt+\left(\curl\boldsymbol{q}-\partial_{t}\boldsymbol{w}\right)_{i}dx^{i}=\ast i_{\mathcal{V}}vol^{4}=\rho dt + \rho v^{i} dx^{i}.\label{wq}
\end{equation}
Here, $\ast$ is the Hodge star operator and we defined:
\begin{subequations}
\begin{align}
\boldsymbol{w}&=\beta_{yz}\partial_{x}+\beta_{zx}\partial_{y}+\beta_{xy}\partial_{z},\\
\boldsymbol{q}&=\beta_{tx}\partial_{x}+\beta_{ty}\partial_{y}+\beta_{tz}\partial_{z}.
\end{align}
\end{subequations}
From equation (\ref{wq}) we see that:
\begin{subequations}
\begin{align}
\rho&=\divv\boldsymbol{w},\label{wq21}\\
\rho\boldsymbol{v}&=\curl\boldsymbol{q}-\partial_{t}\boldsymbol{w}.\label{wq22}
\end{align}
\end{subequations}
The vector fields $\left(\boldsymbol{w},\boldsymbol{q}\right)$ together with the relations \eqref{wq21} and \eqref{wq22} ensure that the 
continuity equation is always satisfied, as one can verify by taking the divergence of \eqref{wq22} and substituting \eqref{wq21} in \eqref{wq22}. 
Note that \eqref{wq21} and \eqref{wq22} are completely analogous to the two Maxwell's equations:
\begin{subequations}
\begin{align}
\frac{n}{\epsilonSI}&=\divv\boldsymbol{E},\label{Mx1}\\  
\muSI\boldsymbol{J}&=\curl\boldsymbol{B}-\muSI\epsilonSI\partial_{t}\boldsymbol{E}.\label{Mx2}
\end{align}
\end{subequations}
with $n$ the charge density, $\epsilonSI$ the vacuum permittivity, $\boldsymbol{E}=-\partial_{t}\boldsymbol{A}-\nabla\phi$ the electric field, $\boldsymbol{B}=\curl\boldsymbol{A}$ the magnetic field, and $\boldsymbol{J}$ the
current density.
Therefore, the independent fields describing $\boldsymbol{v}$ and $\rho$ that will be exploited in the variation of the action functional will be $\boldsymbol{w}$ and $\boldsymbol{q}$.

\subsection{Lagrangian density and proper flow helicity}

Our next task it to determine the Lagrangian and the relevant constraints for the variational relaxation principle.
Since we are looking for a relaxed state in a fluid system, the Lagrangian will be of the standard form kinetic minus 
potential energy: 
\begin{equation}
\mathcal{L}^{\rm fl}=\int_{\Omega}\lr{\frac{1}{2}\rho v^{2}-\frac{P}{\gamma-1}}\,dV. \label{Lfl}
\end{equation}
Furthermore, 
the global entropy $S_{\Omega}$ of equation \eqref{S}
will act as a constraint for the variational problem.
Its value, dictated by the entropy maximization mechanism relaxing the system from the initial ideal configuration, 
will be a dynamical constant of the relaxed state.

Finally, in order to free $\boldsymbol{v}$, we must guarantee that the associated global topological invariant is preserved.
However, since $\boldsymbol{v}$ is not an independent variable, the flow helicity \eqref{Kfl} written in terms of $\boldsymbol{v}$
is not acceptable for the variational problem, as it mixes the topology of $\boldsymbol{q}$ and $\boldsymbol{w}$ [one can verify that the use of $K^{\rm fl}_{\Omega}$ in the variation leads to an unphysical steady state density profile $\rho=\rho_{0}\exp\left(v^{2}/2\tau\right)$]. 
Mathematically, this fact can be understood by noting that the flow helicity $K^{\rm fl}_{\Omega}$ involves higher order derivatives of $\boldsymbol{v}$ than the kinetic energy in \eqref{Lfl}. 
Therefore, the flow helicity $K^{\rm fl}_{\Omega}$ will dissipate at a faster rate than the kinetic energy and cannot serve
as constraint. On this point, compare the order of derivatives of the vector potential $\boldsymbol{A}$ 
in Taylor relaxation \citep{Taylor74,Taylor86} and see \citet{Yoshida02}. 

By setting $\partial_{t}=0$ in \eqref{wq22} and by using \eqref{wq21}, we see that, as in the electrostatic case, $\boldsymbol{w}=\nabla\zeta$ for some potential $\zeta$. Since $\curl\boldsymbol{w}=0$, there is no helicity associated to $\boldsymbol{w}$ in the steady flow limit. Therefore, we are led to introduce the \emph{proper flow helicity}:    
\begin{equation}
Y_{\Omega}=\frac{1}{2}\int_{\Omega}{\boldsymbol{q}\vdot\curl\boldsymbol{q}}\,dV,\label{properKfl}
\end{equation}
We shall prove later on that the Euler-Lagrange equations 
for the relaxed state are such that $Y_{\Omega}\propto K^{\rm fl}_{\Omega}$, with the proportionality factor given by the Lagrange multiplier associated to $Y_{\Omega}$ in the variation.
This fact justifies the choice of $Y_{\Omega}$ as a constraint, since $K^{\rm fl}_{\Omega}$ is an invariant of the Euler equations
\eqref{continuity} and \eqref{momentum}. Mathematically, note that now $Y_{\Omega}$ is well-posed since it involves lower order derivatives than the kinetic energy in \eqref{Lfl}.
	
The total Lagrangian in terms of the independent fields $\left(\boldsymbol{q},\boldsymbol{w},P\right)$ 
and including the constraints \eqref{S}
and \eqref{properKfl} becomes:
\begin{equation}
\mathcal{L}=
\int_{\Omega}{\left[\frac{\left(\curl\boldsymbol{q}-\partial_{t}\boldsymbol{w}\right)^{2}}{2\divv\boldsymbol{w}}
-\frac{P}{\gamma-1}+
\frac{\mu^{\rm fl}}{2}\boldsymbol{q}\vdot\curl\boldsymbol{q}+\tau\frac{\divv\boldsymbol{w}}{\gamma-1}\log{\left(k\frac{P}{\left(\divv\boldsymbol{w}\right)^{\gamma}}\right)}\right]}\,dV
.\label{wqL}
\end{equation}
In this equation $\mu^{\rm fl}$ and $\tau$ are Lagrange multipliers.

\subsection{Variation}  

Variation of \eqref{wqL} with the condition that all variations in the independent fields vanish on $\partial\Omega$ gives:
\begin{subequations}
\begin{align}
\frac{\delta\mathcal{L}}{\delta\boldsymbol{q}}&=\curl\boldsymbol{v}+\mu^{\rm fl}\curl\boldsymbol{q},\label{Lq}\\
\frac{\delta\mathcal{L}}{\delta\boldsymbol{w}}&=\partial_{t}\boldsymbol{v}+\nabla\left(\frac{v^{2}}{2}-\frac{\tau}{\gamma-1}\log\left(\frac{P}{\left(\nabla\vdot\boldsymbol{w}\right)^{\gamma}}\right)\right),\label{Lw}\\
\frac{\delta\mathcal{L}}{\delta P}&=\frac{1}{\gamma-1}\left(\frac{\tau\rho}{P}-1\right).\label{LP}
\end{align}
\end{subequations}
Setting the variations to zero, we obtain the Euler-Lagrange equations: 
\begin{subequations}\label{RxEuler}
\begin{align}
\curl\boldsymbol{v}&=-\mu^{\rm fl}\left(\rho\boldsymbol{v}+\partial_{t}\boldsymbol{w}\right),\label{Lq2}\\
\partial_{t}\boldsymbol{v}&=-\grad\left(\frac{v^{2}}{2}+\tau\log{\rho}\right),\label{Lw2}\\
P&=\tau\rho.\label{LP2}
\end{align}
\end{subequations}

We conclude this section by showing that the system above is such that $Y_{\Omega}=\lr{\mu^{\rm fl}}^{-2}K^{\rm fl}_{\Omega}$. From \eqref{Lq}, we see that:
\begin{equation}
\boldsymbol{v}=-\mu^{\rm fl}\boldsymbol{q}+\grad\psi,
\end{equation}
for some function $\psi$. Therefore:
\begin{equation}
Y_{\Omega}={\frac{1}{2\left(\mu^{\rm fl}\right)^{2}}\int_{\Omega}\left(\boldsymbol{v}-\nabla\psi\right)\vdot\curl\boldsymbol{v}}\,dV=\frac{K^{\rm fl}_{\Omega}}{\left(\mu^{\rm fl}\right)^{2}}-\frac{1}{2\left(\mu^{\rm fl}\right)^{2}}\int_{\Omega}\nabla\vdot\left(\psi\curl\boldsymbol{v}\right)\,dV.\label{YandK}
\end{equation}
If $K^{\rm fl}_{\Omega}$ is constant, $Y_{\Omega}$ is preserved provided that either $\psi=0$ or $\boldsymbol{n}\vdot\curl\boldsymbol{v}=0$ at the boundaries. Note that the latter condition is the same condition required to preserve $K^{\rm fl}$ in the Euler equations \citep[see][]{Moffatt69}. Finally,
let us verify that 
equations \eqref{RxEuler} 
 guarantee that $\frac{dY_{\Omega}}{dt}=0$.
First, consider a fixed domain $\Omega$ such that $\bs{\omega}\vdot\bs{n}=0$ on $\partial\Omega$. Using equations \eqref{RxEuler} and \eqref{YandK}, we have:
\begin{equation}
\begin{split}
\frac{dY_{\Omega}}{dt}&
=\frac{1}{2\left(\mu^{\rm fl}\right)^{2}}\int_{\Omega}{\left(\partial_{t}\bs{v}\vdot\bs{\omega}+\bs{v}\cdot{\partial_{t}\bs{\omega}}\right)}\,dV\\
&=-\frac{1}{2\left(\mu^{\rm fl}\right)^{2}}\int_{\partial\Omega}{\left(\frac{v^{2}}{2}+\tau\log\rho\right)\bs{\omega}\vdot\bs{n}}\,dS=0.
\end{split}
\end{equation}
Now consider a domain $\Omega$ comoving with the fluid. Again, using \eqref{RxEuler} and \eqref{YandK}, we have:
\begin{equation}
\begin{split}
\frac{dY_{\Omega}}{dt}&={\frac{1}{2\left(\mu^{\rm fl}\right)^{2}}}\int_{\Omega}{\frac{d}{dt}\left(\frac{\bs{v}\vdot\bs{\omega}}{\rho}\right)}\,\rho\, dV\\
&={\frac{1}{2\left(\mu^{\rm fl}\right)^{2}}}\int_{\Omega}{\left[\partial_{t}\bs{v}\vdot\bs{\omega}+\frac{\bs{v}\vdot\bs{\omega}}{\rho}\divv{\left(\rho\bs{v}\right)}+\rho\bs{v}\vdot\grad{\left(\frac{\bs{v}\vdot\bs{\omega}}{\rho}\right)}\right]}\,dV\\
&={\frac{1}{2\left(\mu^{\rm fl}\right)^{2}}}\int_{\partial\Omega}{\left[-\left(\frac{v^{2}}{2}+\tau\log\rho\right)\bs{\omega}+\left(\bs{v}\vdot\bs{\omega}\right)\bs{v}\right]\vdot\bs{n}}\,dS.
\end{split}
\end{equation}
Therefore, since $\bs{\omega}\vdot\bs{n}=0$ on $\partial\Omega$, the proper helicity in the comoving domain 
is preserved provided that at the boundary either $\bs{v}\vdot\bs{\omega}=0$ or $\bs{v}\vdot\bs{n}=\bs{0}$.

The third scenario is a time dependent domain $\Omega=\Omega\left(t\right)$. 
In this case, preserving $Y_{\Omega}$ requires a time-dependent Lagrange multiplier $\mu^{\rm fl}=\mu^{\rm fl}\left(t\right)$:
\begin{equation}
\frac{dY_{\Omega}}{dt}=\left[-2Y_{\Omega}\partial_{t}\log\mu^{\rm fl}+{\frac{1}{2\left(\mu^{\rm fl}\right)^{2}}}\left(\int_{\Omega}{\partial_{t}\bs{v}\vdot\bs{\omega}}\,dV+\int_{\partial\Omega}{\left(\bs{v}\vdot\bs{\omega}\right)}\left(\bs{v}^{b}\vdot\bs{n}\right)\,dS\right)\right].
\end{equation}
Here $\bs{v}^{b}$ is the velocity of the boundary. The second term on the right-hand side vanishes due to the boundary condition $\bs{\omega}\vdot\bs{n}=0$. Therefore $Y_{\Omega}$ is constant
provided that:
\begin{equation}
\partial_{t}\left(\mu^{\rm fl}\right)^{2}=\frac{1}{2Y_{\Omega}}\int_{\partial\Omega}{\left(\bs{v}\vdot\bs{\omega}\right)}\left(\bs{v}^{b}\vdot\bs{n}\right)\,dS.
\end{equation}

\section{Relaxed Euler fluid}\label{sec:6}

In this section we discuss the physical implications of the relaxed Euler equations \eqref{RxEuler} 
and the relation of the present model with the Euler equations \eqref{continuity} and \eqref{momentum}. 
First, consider the steady flow limit $\partial_{t}=0$ of system \eqref{RxEuler}: 
\begin{subequations}\label{SteadyRxEuler}
\begin{align}
\curl\boldsymbol{v}&=-\mu^{\rm fl}\rho\boldsymbol{v},\label{StLq}\\
\rho&=\rho_{0}\exp\left(-\frac{v^{2}}{2\tau}\right),\label{StLw}\\
P&=\tau\rho.
\end{align}
\end{subequations}
Equations \eqref{StLq} and \eqref{StLw} correctly reproduce the steady flow solution \eqref{eq:vBeltramiNL} derived in section \ref{sec:3}, with $\alpha_{0}=-\mu^{\rm fl}\rho_{0}$. Furthermore, a fast current makes the system less prone to sustain eddies since
$\curl\boldsymbol{v}$ becomes progressively smaller for growing $v^{2}/\tau$.
This behavior appears to be consistent with scenarios described in \citet{Moffatt14,Hussain86,Rogers87,Wallace92,Scheeler14} where 
vortices 
are progressively eroded by current sheets surrounding them through the Kelvin-Helmholtz instability, while Beltrami states (representing the coherent central part of the eddies) slow down the Kolmogorov cascade toward smaller scales that is responsible for kinetic energy dissipation. The density factor in \eqref{StLq} also suggests that coherent structures 
may be easier to detect in actual experiments if currents do not dominate the turbulent flow and the temperature is high, so that $v^{2}/\tau$ is sufficiently small and the Beltrami configurations can survive for longer times. 
Given $\mu^{\rm fl}$, the Lagrange multiplier for the proper helicity constraint, the model predicts exponential reduction in the ratio of vorticity to velocity as $v^{2}/\tau$ increases, where $\tau$ is the square of the isothermal sound speed and $\mu^{\rm fl}$. 
While the supersonic flow case appears formally as a possibility, it is of dubious physical relevance: (a) supersonic flows are unstable to shock formation and (b) the relaxation timescale must be longer than typical sound transit times across $\Omega$, in order that all parts of the volume can participate in the relaxation process; a supersonic fluid element would thus see changes faster than the relaxation timescale, presumably making relaxation theory inapplicable.
In the strongly subsonic limit, $v^{2}/\tau\rightarrow 0$, the density converges to the constant value $\rho_{0}$ and the flow Beltrami state is linear, with constant proportionality factor between vorticity and velocity . 


By substituting \eqref{Lq2} and \eqref{Lw2} in \eqref{continuity} and \eqref{momentum}, we obtain
the condition that must be satisfied for the relaxed Euler equations 
\eqref{RxEuler} to be a subset of solutions 
of the Euler equations \eqref{Euler}: 
\begin{equation}
\boldsymbol{v}\cross\partial_{t}\boldsymbol{w}=0.
\end{equation}
From the equation above we see that the generalized Beltrami state \eqref{Lq2} 
is a solution of the Euler equations 
if the
displacement $\partial_{t}\boldsymbol{w}$ is aligned with $\boldsymbol{v}$.
Thus, non-aligned displacements, which represent configurations of physical interest as they enable misalignment
between $\boldsymbol{v}$ and $\boldsymbol{\omega}$, may result from  relaxation of a broader class
of fluid equations that include higher order and non-ideal effects such as viscosity. 

\section{Stability analysis}\label{sec:7}

Now we study the stability of the stationary solution of the relaxed Euler fluid. Such a solution is stable against perturbations in an appropriate neighborhood of the stationary state provided that we can show that it is a local minimum of the energy functional $\mc{H}$ of equation \eqref{Energy}.
Let $\left(\bs{v}_{s},\rho_{s}\right)$ be the stationary solution. 
Perturb the system in the following manner:
\begin{subequations}
\begin{align}
\bs{v}&=\bs{v}_{s}+\delta^{1}\bs{v}+\delta^{2}\bs{v}+o\left(3\right),\\
\rho&=\rho_{s}+\delta^{1}\rho+\delta^{2}\rho+o\left(3\right).
\end{align}
\end{subequations}
Here the superscripts refer to the order of the perturbations. 
Recalling that $P=\tau\rho$ and substituting in \eqref{Energy}, we have:
\begin{equation}
\begin{split}
\mc{H}=&\mc{H}_{s}+\delta^{1}\mc{H}+\delta^{2}\mc{H}+o\left(3\right)=\\&\int_{\Omega}{\left[\rho_{s}\left(\frac{v_{s}^{2}}{2}+\frac{\tau}{\gamma-1}\right)\right]}\,dV+\int_{\Omega}{\left[
\rho_{s}\bs{v}_{s}\vdot\delta^{1}\bs{v}+\delta^{1}\rho\left(\frac{v_{s}^{2}}{2}+\frac{\tau}{\gamma-1}\right)\right]}\,dV\\&+\int_{\Omega}{\left[\rho_{s}\left(\bs{v}_{s}\vdot\delta^{2}\bs{v}+\frac{\delta^{1}\bs{v}^{2}}{2}\right)
+\delta^{1}\rho\,\bs{v}_{s}\vdot\delta^{1}\bs{v}+\delta^{2}\rho\left(\frac{v_{s}^{2}}{2}+\frac{\tau}{\gamma-1}\right) \right]}\,dV+o\left(3\right).
\end{split}
\end{equation} 
Therefore, showing stability amounts at proving that $\delta^{1}\mc{H}=0$ and $\delta^{2}\mc{H}>0$.
Now, take the divergence of \eqref{Lq2} and the curl of \eqref{Lw2} to obtain:
\begin{subequations}
\begin{align}
\partial_{t}\rho&=-\divv\left(\rho\bs{v}\right),\\
\partial_{t}\bs{\omega}&=0.
\end{align}
\end{subequations}
The first equation tells us that any perturbation in the density of the system must be 
in the form of a divergence; the second equation implies that the vorticity cannot change
and therefore all perturbations in the fluid velocity must be in the form of a gradient.
Specifically, one finds that:
\begin{subequations}
\begin{align}
\partial_{t}\delta^{1}\bs{v}&=-\grad\partial_{t}\phi^{1}=-\grad\left(\bs{v}_{s}\vdot\delta^{1}\bs{v}+\tau\frac{\delta^{1}\rho}{\rho_{s}}\right),\label{dv1}\\
\partial_{t}\delta^{2}\bs{v}&=-\grad\partial_{t}\phi^{2}=-\grad\left[\bs{v}_{s}\vdot\delta^{2}\bs{v}+\frac{\delta^{1}\bs{v}^{2}}{2}+\tau\left(\frac{\delta^{2}\rho}{\rho_{s}}-\frac{\delta^{1}\rho^{2}}{2\rho_{s}^{2}}\right)\right],\\
\partial_{t}\delta^{1}\rho&=-\divv\partial_{t}\bs{u}^{1}=-\divv\left(\delta^{1}\rho\,\bs{v}_{s}+\rho_{s}\delta^{1}\bs{v}\right),\\
\partial_{t}\delta^{2}\rho&=-\divv\partial_{t}\bs{u}^{2}=-\divv\left(\delta^{2}\rho\,\bs{v}_{s}+\delta^{1}\rho\,\delta^{1}\bs{v}+\rho_{s}\delta^{2}\bs{v}\right).
\end{align}
\end{subequations}
Integrating with respect to $t$ and noting that $\delta^{1}\rho\left(0\right)=\delta^{2}\rho\left(0\right)=0$
and $\delta^{1}\bs{v}\left(0\right)=\delta^{2}\bs{v}\left(0\right)=\bs{0}$, we obtain:
\begin{subequations}
\begin{align}
\delta^{1}\bs{v}&=-\grad\left(\phi^{1}-\phi^{1}\left(0\right)\right),\\
\delta^{2}\bs{v}&=-\grad\left(\phi^{2}-\phi^{2}\left(0\right)\right),\\
\delta^{1}\rho&=-\divv\left(\bs{u}^{1}-\bs{u}^{1}\left(0\right)\right),\\
\delta^{2}\rho&=-\divv\left(\bs{u}^{2}-\bs{u}^{2}\left(0\right)\right).
\end{align}
\end{subequations}
To simplify the notation, we choose $\phi^{1}\left(0\right)=\phi^{2}\left(0\right)=0$
and $\bs{u}^{1}\left(0\right)=\bs{u}^{2}\left(0\right)=\bs{0}$.
Consider now the first variation in the energy:
\begin{equation}
\delta^{1}\mc{H}=\int_{\Omega}\left[\frac{\curl\bs{v}_{s}}{\rm\mu^{fl}}\vdot\grad\phi^{1}-\divv\bs{u}^{1}\left(\frac{v_{s}^{2}}{2}+\frac{\tau}{\gamma-1}\right)\right]\,dV.
\end{equation}
Here we used \eqref{Lq2}. Using the boundary conditions $\bs{\omega}_{s}\vdot\bs{n}=0$ and $\bs{u}^{1}=\bs{0}$ on $\partial\Omega$, one can reduce this integral to:
\begin{equation}
\delta^{1}\mc{H}=\int_{\Omega}\left[\bs{u}^{1}\vdot\nabla\left(\frac{v_{s}^{2}}{2}\right)\right]\,dV.
\end{equation}
Therefore, $\delta^{1}\mc{H}$ vanishes for any $\bs{u}^{1}$, provided that $v_{s}^{2}$ is constant. Alternatively, $\delta^{1}\mc{H}$ can be set to zero if density fluctuations are negligible $\bol{u}^{1}=\bol{0}$.
Consider now the second variation in $\mc{H}$. We have:
\begin{equation}
\delta^{2}\mc{H}=\int_{\Omega}{\left[\rho_{s}\frac{\delta^{1}\bs{v}^{2}}{2}+\bs{v}_{s}\vdot\delta^{1}\bs{v}\,\delta^{1}\rho+\frac{v^{2}_{s}}{2}\,\delta^{2}\rho\right]}\,dV.\label{dH2}
\end{equation}
From \eqref{dH2} we see that $\delta^{2}\mc{H}>0$ whenever the term involving $\bs{v}_{s}\vdot\delta^{1}\bs{v}\,\delta^{1}\rho+v^{2}_{s}\,\delta^{2}\rho/2$ is negligible. This is true, for example, in the following scenarios:
\begin{enumerate}[1. ]
\item Density fluctuations are negligible $\bol{u}^{1}=\bol{u}^{2}=\bol{0}$.
\item The fluctuations in density and velocity are uncorrelated so that their product vanish when averaging over space and $v^{2}_{s}$ is constant (this last condition ensures that the third term in \eqref{dH2} can be written as a vanishing surface integral).
\item The product $\bs{v}_{s}\vdot\delta^{1}\bs{v}$ is, on average, negligible and $v^{2}_{s}$ is constant. 
\item The temperature of the fluid is sufficiently low so that $\left\lvert\delta^{1}\rho/\rho_{s}\right\rvert<<\abs{\delta^{1}\bol{v}/v_{s}}$ as well as $\left\lvert\delta^{2}\rho/\rho_{s}\right\rvert<<\abs{\delta^{1}\bol{v}/v_{s}}^2$, and second and third terms in \eqref{dH2} are smaller than the first one.
\end{enumerate} 
If any of such conditions is satisfied, it follows that the Beltrami equilibrium:
\begin{subequations}
\begin{align}
\curl\bs{v}_{s}&=-\rm\mu^{fl}\rho_{s}\bs{v}_{s},\\
\rho_{s}&=\rho_{0}e^{-v^{2}_{s}/2\tau}.
\end{align}
\end{subequations}
is a local minimum of the energy $\mc{H}$ and therefore is locally stable. 
Finally, it is worth noticing that, when $v^{2}_{s}$ is constant,
$\delta^{1}H$ and the third term in equation \eqref{dH2} exactly vanish. On the other hand, $\rho_{s}=\rho_{0}e^{-v^{2}_{s}/2\tau}$
becomes constant itself, and the Beltrami state is now linear.
This fact could explain from the stand point of stability 
the expected prevalence of linear Beltrami states in plasma relaxation. 

\section{Concluding remarks}\label{sec:8}

In the present paper we have investigated states of maximal relaxation in fluid systems.
These highly relaxed configurations are obtained by freeing the constraints that 
link variations in the fluid position to variations in the dependent fields that define the action.
Following the relaxation theory (MRxMHD) developed in \citet{Dewar15} in the context of ideal magnetohydrodynamics,
we constructed a fluid relaxed state, the relaxed Euler equations \eqref{RxEuler}. 
This system is obtained by unbinding fluid velocity
from fluid position, while enforcing conservation of mass (continuity equation \eqref{continuity}) and global topological invariants (entropy \eqref{S} and proper flow helicity \eqref{properKfl}).
Physically, the relaxed Euler state describes the motion of a fluid that has been driven to a maximum entropy state
by non-ideal effects, such as viscosity. These entropy increasing mechanisms are responsible
for the decoupling between fluid velocity and position, since the system progressively forgets about the initial configuration. 

Under proper conditions, the relaxed Euler equations \eqref{RxEuler} are found to be a subset of solutions to the ideal Euler equations \eqref{Euler}. Furthermore, the derived dynamical equations can account for misalignment between fluid velocity and vorticity, and predict 
a reduction in vorticity as flow increases toward sonic speeds.  
The Beltrami equilibrium, which is a stationary solution of the relaxed Euler equations, is shown to be 
stable against perturbations if density fluctuations are negligible.  
The equivalent Hamiltonian formulation of the relaxed Euler equations was also given (see Appendix~\ref{sec:HamForm}),
 and it was shown in Appendix~\ref{sec:OhmLaw} that the level of anomalous e.m.f. due to turbulent dynamo action required to maintain the double relaxed state can be kept very small by making the magnetic and fluid Beltrami states as isomorphous as possible (which can be done exactly in the limit of strongly subsonic flow).  

The present theory admits a straightforward generalization to MRxMHD of \citet{Dewar15}.
Specifically, the relaxed Euler equations \eqref{RxEuler} can be cast in the form of a \emph{double relaxed} 
magnetohydrodynamic system (DRxMHD) by including the magnetic field and the associated magnetic helicity in the
application of the principle of stationary action.
Such DRxMHD state describes a state of higher relaxation than MRxMHD where not only the magnetic field
but also the fluid flow has undergone relaxation.
A possible topic to be investigated is the application of the present relaxation theory to the
study of MRxMHD equilibrium in plasma confinement devices by introducing the effect of fluid flow 
through numerical simulations \citep[see][]{Hudson12}.

From a purely fluid mechanics perspective, experimental measurements would be useful to
determine the flow conditions that enable the Beltrami configuration.  
These measurements could also give insight into the conservation of fluid helicity in non-ideal systems, 
as described in \citet{Scheeler14}.
Nevertheless, as explained in the construction of the variational principle, the newly
introduced proper flow helicity \eqref{properKfl} is a well-posed functional.
This is in contrast with the standard flow helicity \eqref{Kfl}, which is expected
to dissipate faster than the kinetic energy. 
Finally, an important contribution to 
the stability of the piecewise-relaxed plasma states assumed in MRxMHD theory
 may be made by the coupling of the fluid to the magnetic field in the 
multiple current/vortex sheets surrounding the relaxed fluid regions 
which 
may stabilize these interfaces against Kelvin-Helmholtz instability. 

\section{Acknowledgments}

The research of N.S. was supported by JSPS KAKENHI Grant No. 16J01486 
and that of R.L.D. by Australian Research Council grant DP170102606.
N.S. would like to acknowledge useful discussion with Z. Yoshida, P. J. Morrison, H.M. Abdelhamid, and Y. Ohno. 

\begin{appendix}

\section{Hamiltonian formalism of the relaxed Euler fluid}\label{sec:HamForm}

In this appendix we give the Hamiltonian form of the relaxed Euler fluid \eqref{RxEuler}. 
Such formulation may be convenient when the energy is used as target functional instead of the action integral.
Physical information encapsulated in the Poisson operator is also useful to identify the geometrical properties
of the system.

First, let us recall the Hamiltonian form of the Euler equations \eqref{Euler}.
The fluid Hamiltonian (including the global entropy \eqref{S}) is: 
\begin{equation}
H=\rho \frac{v^{2}}{2}+\frac{P}{\gamma-1}
-\frac{\tau\rho}{\gamma-1}\log\left(k\frac{P}{\rho^{\gamma}}\right).
\end{equation}
Here the independent variables are $\left(\rho,\boldsymbol{v}\right)$ 
with $P=P\left(\rho\right)$. Notice that the Hamiltonian $H$ differs 
from the energy $\mc{H}$ of equation \eqref{Energy}.
The Poisson operator is:
\begin{equation}
\mathcal{J}=\begin{bmatrix}0&-\divv\\-\grad&-\rho^{-1}\boldsymbol{\omega}\cross\end{bmatrix}.\label{JEuler}
\end{equation}
\eqref{JEuler} satisfies the requirements of a Poisson operator, and, specifically, 
the Jacobi identity \citep[on this point see][]{Abdelhamid15}.
We recall that the equations of motion are calculated in the following way:
\begin{equation}
\begin{bmatrix}\partial_{t}\rho\\\partial_{t}\bs{v}\end{bmatrix}=
\mathcal{J}\begin{bmatrix}
\frac{\delta H}{\delta \rho}\\\frac{\delta H}{\delta \bs{v}}
\end{bmatrix}
\end{equation}

The Hamiltonian formalism of the relaxed Euler fluid \eqref{RxEuler}
is given by the Hamiltonian (including the global entropy \eqref{S} and the proper flow helicity \eqref{properKfl}):
\begin{equation}
H=\divv\boldsymbol{w} \frac{v^{2}}{2}
+\frac{P}{\gamma-1}-\frac{\mu^{\rm fl}}{2}\boldsymbol{v}\vdot\curl\boldsymbol{v}-\frac{\tau\divv\boldsymbol{w}}{\gamma-1}\log\left(k\frac{P}{\left(\divv\boldsymbol{w}\right)^{\gamma}}\right).
\end{equation}
Here the independent variables are $\left(\boldsymbol{w},\boldsymbol{v}\right)$
 with $P=P\left(\rho\right)$.
The Poisson operator is the 
 symplectic matrix:
\begin{equation}
\mathcal{J}=\begin{bmatrix}0&-1\\1&0\end{bmatrix}.
\end{equation}

\section{Ohm's Law in Relaxed MHD}\label{sec:OhmLaw}

In resistive MHD the electric field $\bs{E}' = \bs{E} + \bs{v}\cross\bs{B}$ in the local rest frame of each fluid element obeys Ohm's Law, $\bs{E}' = \eta \bs{J}$, where $\eta$ is the resistivity and, in SI units, the electric current $\bs{J} = \curl\bs{B}/\upmu_0$, $\bs{B}$ being the magnetic field. Ideal MHD sets $\eta$ to zero, to give the \emph{ideal Ohm's Law}, 
\begin{equation}\label{eq:IOhm}
	\bs{E} + \bs{v}\cross\bs{B} = 0
\end{equation}
within $\Omega$. Assuming, as we do throughout this paper, there are no gaps [see \cite{Dewar78}] in the boundary, ideal MHD obeys \eqref{eq:IOhm} everywhere on $\partial\Omega$ as well. Ideal MHD then enforces trivially the \emph{general} perfectly conducting boundary condition that $\bs{E}'$ have no component tangential to $\partial\Omega$,
\begin{equation}\label{eq:Etgt}
	\bs{n}\cross(\bs{E} + \bs{v}\cross\bs{B}) = 0 \;.
\end{equation}

In ideal MHD \eqref{eq:IOhm} is used in Faraday's Law,
\begin{equation}\label{eq:Faraday}
	\curl\bs{E} = -\frac{\partial\bs{B}}{\partial t} \;,
\end{equation}
to eliminate the electric field, giving the time evolution equation for $\bs{B}$. In the Lagrangian picture it is this equation that leads to the holonomic ``frozen-in flux'' constraint giving variations of $\bs{B}$ in terms of $\bs{\xi}(\bs{x},t)$, the field of fluid displacements away from the Lagrangian mean flow [see e.g. \cite{Dewar70} and \cite{Buhler09}] . 
However, in \emph{relaxed} MHD [\cite{Dewar15}] this infinity of microscopic constraints is replaced by the macroscopic constraint of conserved magnetic helicity over $\Omega$ (except on $\partial\Omega$, where ideal constraints are retained). Then the magnetic field $\bs{B}$ no longer has its own equation of motion but instead obeys the Beltrami equation,
\begin{equation}\label{eq:BeltramiMag}
	\curl\bs{B} = \mu(t)\bs{B}
\end{equation}
under the ideal boundary condition of tangentiality to $\partial\Omega(t)$, and is determined uniquely by giving as many fluxes as are required by the topology of $\Omega$ [\cite{Yoshida_Giga_90}]. Below we shall also use the vector potential representation $\bs{B} = \curl\bs{A}$ and assume Coulomb gauge, $\divv\bs{A} = 0$.

Like $\bs{B}$, in relaxed MHD $\bs{A}$ is completely determined, but, as the ideal Ohm's Law does not apply, $\bs{E}$ is only \emph{partially} determined: \eqref{eq:Faraday} gives its curl but its divergence is unknown as the plasma is only quasineutral.\footnote{If we knew $\rho_{\rm e}$, the electric charge density, we could use Poissson's equation $\divv{\bs{E}} = \rho_{\rm e}/\epsilonSI$, where $\upepsilon_0$ is the permittivity of free space; but $\rho_{\rm e}/\epsilonSI$ is indeterminate in the ordering used in ideal MHD [see e.g. \S 5.3 of \cite{Hosking_Dewar_15}], so $\divv{\bs{E}}$ is unconstrained.} However we can \emph{infer} $\bs{E}$ by first writing it in the potential representation $\bs{E} = -\grad\Phi - \partial\bs{A}/\partial t$, to enforce the Faraday constraint \eqref{eq:Faraday}, and then making $\bs{E}$ agree as closely as possible with the ideal Ohm's law \eqref{eq:IOhm} by \emph{minimizing} the discrepancy functional
\begin{equation}\label{eq:RxOhm}
	\bs{\Delta}[\Phi] \equiv -\grad\Phi - \frac{\partial\bs{A}}{\partial t} + \bs{v}\cross\bs{B} \;.
\end{equation}
This will give a measure of the minimum amount of dynamo effect (assumed to be caused by averaged-over mesoscale turbulence) required to break the ideal Ohm's law. Below we show that the $L^2$ norm leads to a well-posed variational problem:
\begin{equation}\label{eq:Deltamin}
	||\bs{\Delta}||^2 = \min_{\Phi}\int_{\Omega}(\bs{\Delta})^2\, \d V \;. 
\end{equation}
Given a variation $\delta\Phi$, the variation in $(\bs{\Delta})^2$ is $\delta\bs{\Delta} = -2\bs{\Delta}\vdot\grad\delta\Phi$. Confining variations to within $\Omega$ and integrating by parts we find the corresponding Euler--Lagrange equation $\divv\bs{\Delta} = 0$, giving a Poisson equation for $\Phi$,
\begin{equation}\label{eq:PhiEL}
\begin{split}
	\nabla^2\Phi  &= \divv\left(\bs{v}\cross\bs{B} - \frac{\partial\bs{A}}{\partial t}\right) \\
				&=  \bs{B}\vdot\curl\bs{v} - \bs{v}\vdot\curl\bs{B} 
\end{split}
\end{equation}

If we were to require also that $||\bs{\Delta}||$ be minimal with respect to localized variations $\delta\Phi$ on $\partial\Omega$ we would obtain, as the surface Euler--Lagrange equation, the natural boundary condition
$\bs{n}\vdot\bs{\Delta} = 0 \:\:\text{on}\:\partial\Omega$, which would be equivalent to requiring $\bs{n}\vdot(\bs{E} + \bs{v}\cross\bs{B}) = 0$. However this is in general incompatible with the physically correct boundary condition \eqref{eq:Etgt}, so we cannot require the Ohm's law discrepancy to be minimal with respect to \emph{localized} surface variations in potential. Below we use surface variations compatible with \eqref{eq:Etgt} to allow further minimization and thus to complete the inference of $\bs{E}$.

For simplicity, henceforth assume the boundaries to be fixed and the velocity field $\bs{v}$ to be the steady relaxed flow given by the nonlinear Beltrami equation \eqref{eq:vBeltrami}, with $\alpha(\bs{v})$ given by \eqref{eq:BeltramiBernoulli}. Equation (\ref{eq:PhiEL}) then becomes
\begin{equation}\label{eq:PhiELsteady}
	\nabla^2\Phi  =  (\alpha - \mu)\,\bs{v}\vdot\bs{B} \;,
\end{equation}
which is a Poisson equation with charge density 
\begin{equation}\label{eq:charge}
	\rho_{\rm e} = \epsilonSI(\mu-\alpha)\,\bs{v}\vdot\bs{B} \;.
\end{equation} 
For very subsonic flow we can make the charge density very small, $O(\mathrm{M}^2)$, where M is the Mach number $v_{\rm max}/C_{\rm s}$, by choosing initial conditions such that $\mu = \alpha_0$. Because $\alpha = \alpha_0[1 + O(\mathrm{M}^2)]$, the fields $\bs{v}$ and $\bs{B}$ then obey almost the same Beltrami equations, so the near cancellation between $\alpha$ and $\mu$ will persist even if the boundary is deformed adiabatically away from its initial shape. However, as noted in the footnote to this Appendix, $\divv\bs{E}$ does not need to be small in ideal MHD, so \eqref{eq:charge} creates no conflict with ideal MHD even if $\mu \neq \alpha_0$ or M is $O(1)$.

With fixed boundaries, both $\bs{B}$ and $\bs{v}$ are tangential to $\partial\Omega$, so $\bs{v}\cross\bs{B}$ is purely normal. Then \eqref{eq:Etgt} implies $\bs{n}\cross\grad\Phi = 0$ on $\partial\Omega$, or, equivalently, the Dirichlet boundary conditions $\Phi = V_{k} = \const$ on each electrically isolated component, $\partial\Omega_{k}$, of $\partial\Omega = \cup_k\partial\Omega_{k}$, these component subsurfaces being disjoint: $\partial\Omega_{k}\cap\partial\Omega_{l} = \emptyset$ for $k\neq l$.

Physically, this is the well-known problem of solving Poisson's equation in a region enclosed by conducting surfaces at potentials $V_{k}$. For example, the inner and outer tori bounding an annular toroid may be at different potentials, forming a kind of capacitor or battery. To infer these potentials $V_{k}$ we minimize the discrepancy $\Delta$ in a similar way to that used to find the natural boundary condition, but with $\delta\Phi = \delta V_{k}$ over the \emph{whole} of $\partial\Omega_k$, giving a sufficient set of stationarity conditions $\int_{\Omega_k}\bs{n}\vdot\bs{\Delta}\,\d S = 0$ to determine the $V_{k}$ up to an arbitrary overall constant.

These conditions mean that the ideal Ohm's law is satisfied \emph{on average} on each isolated subsurface. In terms of the potential $\Phi$ we can write these conditions as an appended set of averaged Neumann boundary conditions,
\begin{equation}\label{eq:surfPhiEL}
	\int_{\partial\Omega_k}\bs{n}\vdot\grad\Phi\,\d S = \int_{\partial\Omega_k}\bs{n}\vdot\bs{v}\cross\bs{B}\,\d S \;.
\end{equation}

It is important to ask whether the minimum $||\bs{\Delta}||$ can be zero in steady state --- if it were, the ideal Ohm's law could be satisfied exactly, eliminating the need to invoke an ongoing dynamo effect. Unfortunately it appears this cannot be done in general, as it would mean solving the equation $\grad\Phi = \bs{v}\cross\bs{B}$ (setting $\pa\bs{A}/\partial t = 0$ in steady state). This has the solvability condition $\curl(\bs{v}\cross\bs{B}) = 0$, which there is no reason to suppose is true in general.

Assuming $\Phi$ satisfies \eqref{eq:PhiEL}, integrate by parts in \eqref{eq:Deltamin} to give
\begin{equation}\label{eq:DeltaminAlt}
\begin{split}
	||\bs{\Delta}||^2 &= \int_{\Omega} (-\grad\Phi + \bs{v}\cross\bs{B})\vdot\bs{\Delta}\, \d V\\
				     &= \int_{\Omega} \left[-\divv(\Phi\bs{\Delta}) + \Phi\divv\bs{\Delta} + \bs{v}\cross\bs{B}\vdot\bs{\Delta}\right]\, \d V  \\
				     &= \int_{\Omega} \bs{v}\cross\bs{B}\vdot\bs{\Delta}\, \d V  
\;, 
\end{split}
\end{equation}
where we used \eqref{eq:PhiEL} in the form $\divv\bs{\Delta} = 0$ and Gauss' theorem with \eqref{eq:surfPhiEL} in the form $\int_{\Omega_k}\bs{n}\vdot\bs{\Delta}\,\d S = 0$ to eliminate the first two terms on the second line.\\

\end{appendix}

\end{document}